\documentclass[aps,prl,twocolumn,superscriptaddress,letterpaper,floatfix,nofootinbib]{revtex4-2}
\usepackage[english]{babel}
\usepackage[T1]{fontenc}
\usepackage[colorlinks=true,linkcolor=blue,citecolor=blue,urlcolor=blue]{hyperref}

\usepackage{graphicx}
\usepackage{bm}
\usepackage{amsmath,amssymb}
\usepackage{epstopdf}
\usepackage{color}
\usepackage{isotope}
\usepackage{bbding}
\usepackage{siunitx}
\usepackage{orcidlink}

\usepackage{comment}
\usepackage[switch]{lineno}
\usepackage[capitalize]{cleveref}

\usepackage{multirow,booktabs,dcolumn}
\renewcommand*{\arraystretch}{1.3}
\definecolor{dgreen}{rgb}{0.0,0.5,0.0}
\usepackage[normalem]{ulem}

\begin{document}

\title{Candidate Toroidal Electric Dipole Mode in the Spherical Nucleus $^{58}$Ni}

\newcommand{\TUD}{Institut f\"ur Kernphysik, Technische Universit\"at Darmstadt,
D-64289 Darmstadt, Germany}

\newcommand{\JINR}{ Laboratory of Theoretical Physics, Joint Institute for
Nuclear Research, Dubna, Moscow region, 141980, Russia}

\newcommand{\DSU}{State University "Dubna", Dubna, Moscow region, 141980, Russia}

\newcommand{\RCNP}{Research Center for Nuclear Physics, Osaka University,
Ibaraki, Osaka 567-0047, Japan}

\newcommand{\Erlangen}{Institut f\"ur Theoretische Physik II, Universit\"at
Erlangen, D-91058, Erlangen, Germany}

\newcommand{\Almaty}{Institute of Nuclear Physics Almaty, Almaty Region, Kazakhstan}

\newcommand{\FHU}{Faculty of Radiological Technology, 
Fujita Health University, Aichi 470-1192, Japan}

\newcommand{\UWS}{School of Computing, Engineering, and Physical Sciences,
University of the West of Scotland, Paisley PA1 2BE, United Kingdom}

\newcommand{\SUPA}{Scottish Universities Physics Alliance, United Kingdom}

\newcommand{\Praha}{Institute of Particle and Nuclear Physics, Charles University,
CZ-18000, Praha 8, Czech Republic}

\newcommand{\Bratislava}{Institute of Physics, Slovak Academy of Sciences,
84511, Bratislava, Slovakia}

\def\Label{}

\author{P.~von~Neumann-Cosel}
\email[Contact Peter von Neumann-Cosel\ ]{(vnc@ikp.tu-darmstadt.de)}
\affiliation{\TUD}

\author{V.~O.~Nesterenko}
\email[Contact Valentin Nesterenko \ ]{(nester@theor.jinr.ru)}
\affiliation{\JINR}
\affiliation{\DSU}

\author{I.~Brandherm}
\affiliation{\TUD}

\author{P.~I.~Vishnevskiy}
\affiliation{\JINR}
\affiliation{\Almaty}

\author{P.-G.~Reinhard}
\affiliation{\Erlangen}

\author{J.~Kvasil}
\affiliation{\Praha}

\author{H.~Matsubara}
\affiliation{\RCNP}
\affiliation{\FHU}

\author{A.~Repko}
\affiliation{\Bratislava}

\author{A.~Richter}
\affiliation{\TUD}

\author{M.~Scheck}
\affiliation{\UWS}\affiliation{\SUPA}


\author{A.~Tamii}
\affiliation{\RCNP}

\date{\today}


\begin{abstract}

Dipole toroidal modes appear in many fields of physics. 
In nuclei, such a mode was predicted more than 50 years ago, but clear experimental evidence was lacking so far. 
Using a combination of high-resolution inelastic scattering experiments with photons, electrons and protons, we identify for the first time candidates for toroidal dipole excitations in the nucleus $^{58}$Ni and demonstrate that transverse electron scattering form factors represent a relevant experimental observable to prove their nature. 
\end{abstract}

\maketitle

\begin{figure}[t] 
\centering
\includegraphics[width=1.0\columnwidth]{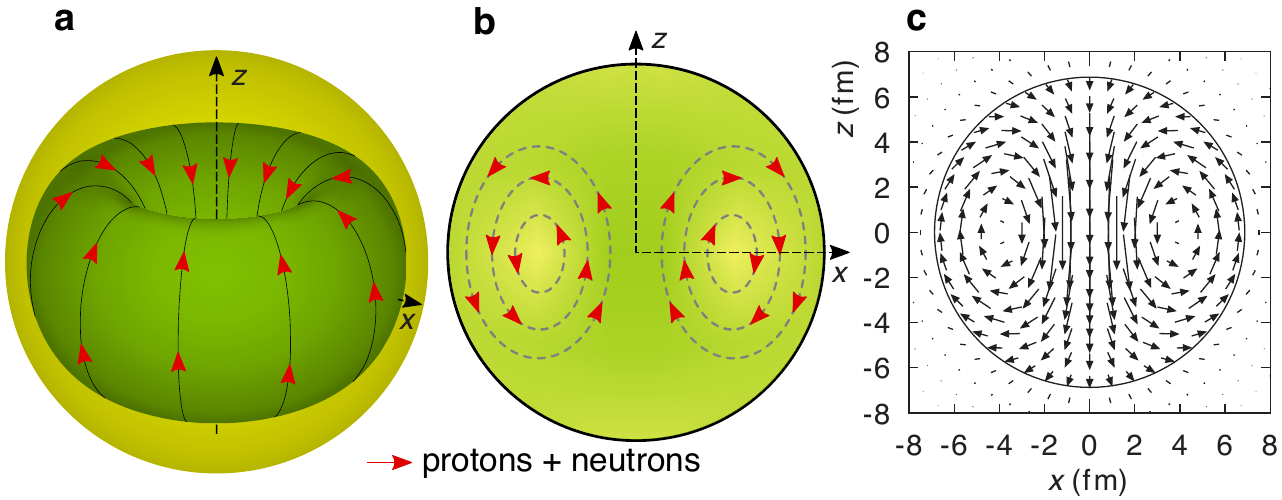}
\caption{Nuclear toroidal excitations. (a) Schematic view and  (b) its cut in the $x$-$z$ plane.
(c): Same as (b) for the toroidal mode predicted in the nucleus $^{208}$Pb \cite{rep13}.
The arrows mark the current along stream lines and their length is a measure of the current density.}
\label{fig1}
\end{figure}

{\it Introduction} -- Toroidal modes appear in a wide variety of fields ranging from solid-state physics \cite{dub90},
metamaterials \cite{kae10} and metaphotonics \cite{ara20,che24} to heavy-ion collisions \cite{iva23} and anapole dark matter \cite{anapole}. Its simplest form is an electric toroidal dipole where the current is circulating on the surface of a torus \cite{Dub74,sem81,Nanz16} as illustrated in Figs.~\ref{fig1}(a,b). Maybe the oldest example is Hill's spherical vortex in hydrodynamics, predicted in 1894  as a stationary solution of the Euler equations for an incompressible fluid
\cite{hil94}.  It appears in turbulent flow for a variety of classical fluids and gases. Actually, each reader of this paper permanently produces (invisible) Hill's vortex rings  in the surrounding air when breathing out. 
Smoke rings are a way to make them visible.

This article aims to identify clear signals for nuclear vorticity in terms of toroidal flow in electric dipole modes at low excitation energy, called toroidal dipole resonance (TDR). The nuclear TDR is a specific example realized in a finite quantum system with a strong interaction \cite{sem81,paa07,rep19}, which makes it a unique study object.
At variance with classical systems where the toroidal vorticity is produced by turbulence, the nuclear TDR is a quantum effect originating from the nuclear shell structure.
In Landau theory of quantum fluids, this resonance belongs to the class of "zero sound" modes with elastic properties \cite{bas93}.  Panel \ref{fig1}(c) shows, as a realistic example, the nuclear current distribution of the TDR predicted in the nucleus $^{208}$Pb \cite{rep13}. Here, the stream lines resemble irregular ellipses. They fill up all the volume of the nucleus, with a strong central flow along the $z$ axis (which can be defined in an experiment by the beam direction).  Note that, unlike a classical Hill’s vortex, in case of the nuclear TDR the nucleons do not exhibit full circulations but small oscillations along the stream lines on the toroidal surface.

First attempts to identify the TDR were made using inelastic $\alpha$-scattering reaction, see e.g.~Ref.~\cite{uch04}.
However, despite the general interest and theoretical predictions ranging back more than 50 years, no strong experimental evidence of the nuclear TDR has been reported so far \cite{sav13,bra19,lan23,rep19}. Here we investigate candidates for toroidal dipole states (angular momentum and parity $J^{\pi}=1^-$) in the nucleus $^{58}$Ni, which can be identified from the combined analysis of a set of high-resolution inelastic scattering experiments with photons, electrons and protons now available
\cite{met84,met87,bau00,shi24,bra24} with emphasis on the $(e,e')$ reaction \cite{met84,met87}. 
The theoretical analysis is performed with density functional theory using Skyrme functionals \cite{ben03}.

{\it General properties of the toroidal mode} --
The toroidal part naturally appears in the expansion of the nuclear current transition density $\delta \mathbf{j}(\mathbf{r})$. 
For this aim, it is instructive to  use the Chandrasekhar-Moffat
decomposition \cite{cha61,Moffat,dub90,Nanz16}
\begin{equation}
 \label{eq:dj}
 \delta\mathbf{j}(\mathbf{r}) = \mathbf{\nabla} \phi(\mathbf{r}) + \mathbf{\nabla}\times (\mathbf{r}\psi(\mathbf{r})) +
 \mathbf{\nabla}\times \mathbf{\nabla}\times(\mathbf{r} \chi(\mathbf{r})) ,
\end{equation}
where $\phi, \psi, \chi$ are some scalar coordinate-dependent functions (often called Debye potentials).
As compared with the familiar Helmholtz expansion, this decomposition allows to separate magnetic $\mathbf{\nabla}\times (\mathbf{r}\psi(\mathbf{r}))$ and electric $\mathbf{\nabla}\times \mathbf{\nabla}\times(\mathbf{r} \chi(\mathbf{r}))$ vortical terms. The last double-curl term produces (after  extraction of its long-wave fraction using Siegert's theorem) the toroidal flow \cite{kva11}. Equation~(\ref{eq:dj}) is also known as Neumann-Debye representation \cite{gray77}.

Further, the toroidal transition operator appears as second-order term in the electric multipole operator \cite{BMv1}
\begin{eqnarray}\label{ME_oper1}
&&\hat{M}(E\lambda\mu, q) =
-i \frac{(2 \lambda + 1)!!}{c q^{\lambda+1} (\lambda+1)}
\\
&\cdot&\int\!d^3r
\hat{\mathbf{j}}_\text{nc}(\mathbf{r})\!\cdot\!
[\mathbf{\nabla}\!\times\!(\mathbf{r}
\!\times\!\mathbf{\nabla})j_{\lambda}(qr) Y_{\lambda \mu}(\hat{\mathbf{r}})],
\nonumber
\end{eqnarray}
where $q$ denotes the momentum transfer, $\hat{\mathbf{j}}_{\text nc}(\mathbf{r})$ the operator of the nuclear current \cite{BMv1}, $j_{\lambda}(qr)$ a spherical Bessel function,  and $Y_{\lambda \mu}(\hat{\mathbf{r}})$ a spherical harmonics. 
In the  long-wavelength approximation ($q\rightarrow 0$),
we can restrict ourselves to the first two terms in the Taylor expansion of the Bessel function and get\cite{kva11}


\begin{equation}
\label{E+ktor}
\hat{M}(E\lambda\mu, q) \approx \hat{M}(E\lambda\mu)
+ q\:\hat{M}_{\text{tor}}(E\lambda\mu) \;
\end{equation}
with the standard  electric operator  $\hat{M}(E\lambda\mu) \propto r^\lambda Y_{\lambda\mu}$ and the second-order toroidal operator $\hat{M}_{\text{tor}}(E\lambda\mu)$ \cite{kva11} 
(see details 
in the Supplemental Material (SM) \cite{SM}).
Note that the toroidal operator includes
$\mathbf{\nabla} \times \hat{\mathbf j}_{\text nc}$ 
and thus generates basically vortical transitions.

To experimentally observe the second-order toroidal mode, we need to suppress the dominant $E1$ transitions, 
which is generally a non-trivial problem. 
For example, in  metamaterials this is achieved only by a special construction of the system \cite{kae10}.
Atomic nuclei have a principle advantage with the possibility to use selective reactions and resolve individual states  \cite{har01,nes18,Ne_PRC19}. 
First, as shown below, toroidal modes dominate over irrotational modes at backward angles in electron inelastic scattering.  
Second,  the dominant $E1$ transitions in nuclei are mainly of isovector (IV) nature, where protons and neutrons move in opposite directions. 
Their impact can be suppressed by investigating dominantly isoscalar (IS) transitions with in-phase motion of protons and neutrons.

The reduced $B(E1)$ transition strength related to the first order dipole operator $\propto rY_{1\mu}$ is irrotational. Its IV fraction is concentrated in the giant dipole resonance (GDR). The IS dipole strength $B(IS1)$ for this operator is associated with the spurious center-of-mass motion of the entire nucleus and thus is not a part of the intrinsic dipole excitations. The second order transition dipole operators $\propto r^3Y_{1\mu}$ include the vortical toroidal operator and its irrotational counterpart, the compression mode forming the ISGDR, see \cite{kva11} and SM \cite{SM}. The toroidal and compression modes can compete.  Thus, we need to consider both of them. Previous studies showed that the low-energy part of the toroidal mode is basically of IS nature \cite{rep19}. 

The TDR, as a concentration of the dipole toroidal strength at energies  $E_{\rm x} \approx (40-60)A^{-1/3}$ MeV, was predicted by a variety of nuclear models: hydro/fluid-dynamical \cite{bas93,mis06}, quasiparticle-phonon
\cite{rye02}, relativistic \cite{vre02}, and non-relativistic \cite{kva11,rep13,rep19} Quasiparticle Random-Phase Approximation (QRPA). 
The high-energy part of the TDR is concealed by the GDR located at $E_{\rm x} \approx 70-90A^{-1/3}$ MeV, 
but low-energy toroidal strength is accessible to experimental investigation \cite{rep19}. 
Further, individual low-energy toroidal states were predicted in light nuclei within Skyrme QRPA \cite{nes18} and Antisymmetrized Molecular Dynamics (see, e.g., Ref.~ \cite{kan18a}).

\begin{figure*} 
\centering
\includegraphics[width=0.9\linewidth]{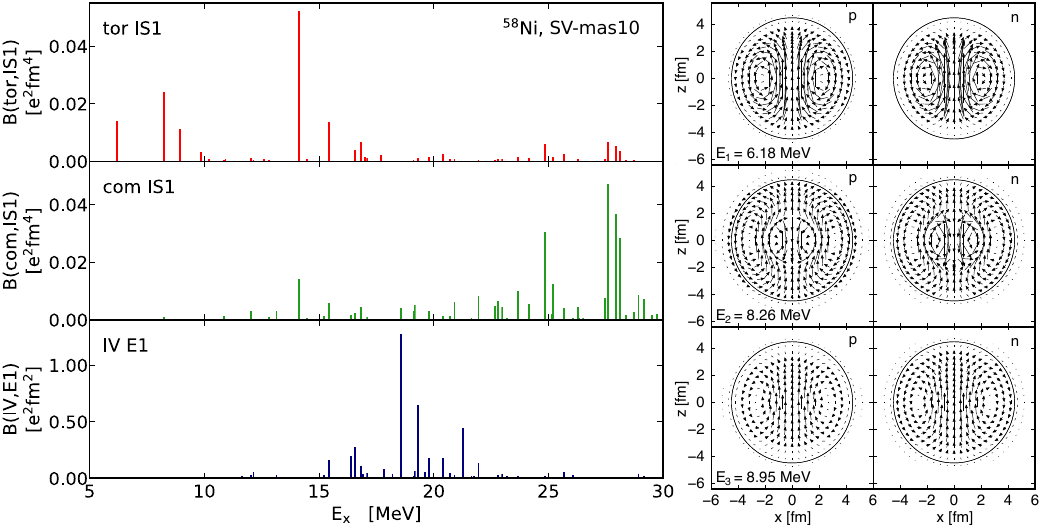}
\caption{
Left: IS1 Toroidal (top), IS1 compression (middle) and IV E1 (bottom) strengths in $^{58}$Ni calculated with the DFT approach described in the text and the SV-mas10 interaction.
Right: Proton and neutron nuclear currents for the three lowest toroidal transitions.}
\label{fig2}
\end{figure*}

{\it Model predictions}
-- Calculations are performed within nuclear density-functional theory (DFT) using the Skyrme functional \cite{ben03}. 
Small-amplitude dipole oscillations are treated within fully self-consistent QRPA \cite{rep15} 
using the parametrization SV-mas10 \cite{klu09}, which gives the most reasonable description of low-energy dipole spectra in $^{58}$Ni 
(for details see the SM \cite{SM}).

The left part of Fig.~\ref{fig2} shows the reduced transition probabilities in $^{58}$Ni predicted
for the three dipole strengths defined above. 
The isovector $B(E1)$ strength is mainly located in the
GDR region at 15-22 MeV, while the IS dipole compression strength $B$(com,IS1) is concentrated at $24-30$ MeV, where it forms the ISGDR \cite{har01}.
The $B$(tor,IS1) strength is broadly distributed over the whole energy range, but its most significant part is found at $5-15$ MeV.

For our aims, the low-energy toroidal part, not concealed by the GDR or the ISGDR, is most interesting.
The upper panel of Fig.~\ref{fig2} shows that three lowest dipole states predicted at 6.18, 8.26 and 8.95 MeV have considerable toroidal strength. 
Their energies match well the energies  6.03, 8.24 and 8.87 MeV of low-energy $1^-$ states prominently seen in different reactions \cite{nes10} and, in particular, in an $(\alpha,\alpha^\prime\gamma)$ experiment selective to IS strength \cite{poe92}. 
The distributions and mutual similarity of proton and neutron currents for these three  states, displayed in the right part of Fig.~\ref{fig2}, demonstrate a clear toroidal flow with IS character, corresponding to Fig.~\ref{fig1}. 
Figure~\ref{fig2} also shows that the IS1 toroidal strength strongly dominates over the compression one at low excitation energies.
Thus, the considerable low-energy IS1 strength observed in $^{58}$Ni \cite{poe92} should originate from the toroidal rather than the compression mode.
Note that significant IS1 strength observed in medium and heavy nuclei at 10-20 MeV \cite{uch04} including $^{58}$Ni \cite{lui06} was also suggested being toroidal \cite{vre02}.

As pointed out above, the nuclear TDR has a quantum mean-field origin.
It is known that the strongest electric dipole two-quasiparticle (2qp) transitions are those from the valence to the next quantum shell \cite{har01,rin80} and their energies coincide with the TDR region. 
These pure 2qp $E1$ excitations 
have a mixed irrotational and vortical character. 
As demonstrated in the SM \cite{SM}, on the way from mean field to QRPA, the IV dipole residual interaction up-shifts most of the irrotational $E1$ strength to produce the GDR, while the vortical strength remains in its original low-energy range 
and forms the TDR.

{\it Experimental evidence} --
The experimental identification of possible toroidal excitations in $^{58}$Ni is based on a unique set of data from high-resolution inelastic scattering experiments with photons \cite{bau00,shi24}, electrons \cite{met87}, and protons \cite{bra24} permitting a state-by-state analysis. 
All experiments have been performed in kinematics where dipole excitations are strongly favored but require a separation of $E1$ and $M1$ transitions.

The ($p,p^\prime$) experiment was executed at the Research Center for Nuclear Physics, Osaka University, Japan.
Selectivity to dipole excitations can be achieved in measurements at very forward angles close to
$0^\circ$ with respect to the incident beam direction.
The multipole decomposition analysis of $E1$ and $M1$ contributions to the cross sections and the conversion to equivalent electromagnetic transitions strengths are described in Ref.~\cite{vnc19} and the application to the $^{58}$Ni experiment in Ref.~\cite{bra24}.
The  ($\gamma,\gamma^\prime$) reaction is a selective probe for dipole excitations and a distinction of their electric or magnetic character is possible using a beam of linearly polarised photons \cite{zil22}.

The inelastic electron scattering ($e,e^\prime$) experiment was performed  at low momentum transfer and backward angles~\cite{met87}.
As discussed below, the measured cross sections contain longitudinal and transverse parts, whose ratio varies at a given electron energy as a function of the scattering angle. Magnetic transitions are enhanced at backward angles.
Thus, all dipole transitions in $^{58}$Ni excited at the largest angles were initially assigned as $M1$~\cite{met87}.
A possible $E1$ character was not considered since transverse contributions are expected to be very small for the isoscalar compression and isovector modes due to their irrotational nature.
However, the combined analysis of new $(p,p^\prime)$ and ($\gamma,\gamma^\prime$) experiments described in Ref.~\cite{bra24} proves that the dipole excitations of our interest
are $E1$.
Here, we suggest a new interpretation of these transitions as part of the toroidal mode.

The ($e,e^\prime$) data have been measured at an approximately constant incident electron energy
$E_i \simeq 50$ MeV \cite{met87}. In Plane-Wave Born Approximation
(PWBA), the inelastic electron scattering cross section for $E\lambda$ transitions reads \cite{HB83}
\begin{equation}
\label{eq:eeprime}
\frac{\sigma}{\sigma_\mathrm{Mott}} (\theta)\propto
|F^C_{E\lambda}(\theta)|^2
+ \left[1/2+\tan^2 (\theta/2)\right]
|F^T_{E\lambda}(\theta)|^2,
\end{equation}
where $\sigma_\mathrm{Mott}$ is the corresponding elastic cross section from scattering off a pointlike object with the same charge, and $F^{C/T}$ denote the  Coulomb (or longitudinal) and transverse form factors resulting from the interaction of the electron with the nuclear charge and current distributions, respectively 
(for details see \cite{SM}).
The transverse form factor is expected to be sensitive to the toroidal flow.
\begin{figure}[t] 
\centering
\includegraphics[width=0.85\columnwidth]{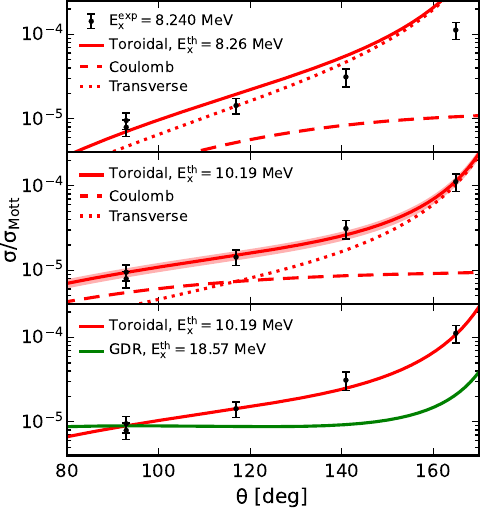}
\caption{Electron scattering cross sections 
of the toroidal candidate at 8.240 MeV \cite{met87} compared to QRPA predictions using the SV-mas10 interaction.
Top: Prominent theoretical toroidal excitation at 8.26 MeV.
Middle: Best fit result obtained as described in the text.
Bottom: Strongest theoretical isovector E1 transition (green line). The theoretical curves are normalized to the data at the most forward angle.
}	
\label{fig4}
\end{figure}

As an example, we explore in Fig.~\ref{fig4} the transition at 8.240 MeV prominently seen in all the considered reactions.
The experimental ($e,e^\prime$) cross sections \cite{met87} are shown as a function of scattering angle. 
In the upper part, the experimental data are compared with the QRPA results for the $1^-$ state at 8.26 MeV closest in energy, which, following Fig.~\ref{fig2}, is one of main toroidal candidates.
Clearly, the transverse contribution determines the behavior of the cross section, especially at large angles. 
Some discrepancy with the data can be explained by the limitations of QRPA calculations omitting, e.g., the coupling to complex configurations, which can modify the structure of the states \cite{bau00} and 
redistribute the toroidal strength between the states. 
Besides, systematic model uncertainties due to the restriction to Skyrme forces can also influence the results~\cite{SM}.

In the middle panel of Fig.~\ref{fig4}, the experimental data are compared with the best-fit QRPA result
described in the SM \cite{SM}. 
It provides an excellent description of the data.
Finally, in the lower panel, the angular dependence for the toroidal and GDR cross sections are compared. 
The latter is represented by the most collective GDR state at 18.6 MeV. 
For comparison, the theoretical cross sections are normalized to the data at the smallest scattering angle measured. 
Clearly, the GDR does not show the necessary strong slope of  $\sigma/\sigma_\mathrm{Mott}$ towards larger scattering angles which complies with its dominantly irrotational nature. 
Note that, following Eq.~(\ref{eq:dj}), the GDR irrotational nuclear current has the form $\delta {\bf j} \propto \nabla \phi$ and so its contribution  $\nabla \times \delta {\bf j} $ to the transverse electric form factor is much suppressed \cite{SM}. 
The same argument should hold for the irrotational compression mode. 
Furthermore, following Fig.~\ref{fig2}, the compression mode strength is negligible at low excitation energies.
Hence, our analysis signifies the toroidal nature of low-energy $1^-$ states in $^{58}$Ni.
The dominance of the low-energy toroidal strength over the compression one is predicted to be a common feature across the nuclear chart \cite{rep19,rep13}, although states with significant compression fraction can appear \cite{rep19,nes18}.

\begin{figure} 
\centering
\includegraphics[width=\columnwidth]{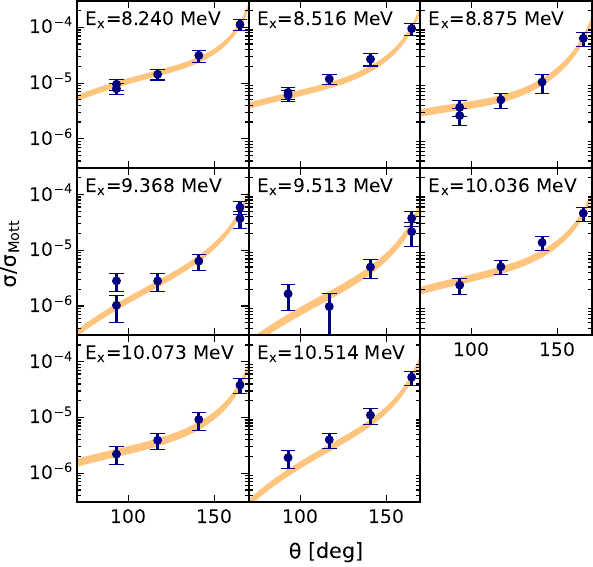}
\caption{
Electron scattering cross sections (blue circles) 
of all toroidal candidates \cite{met87} compared with the best results from $\chi^2$ fits of all theoretical $E1$ excitations below 11 MeV (orange bands).
The energies of the QRPA states and the normalization factors $N$ are quoted in the figure.}
\label{fig:compall}
\end{figure}

The electron scattering data for all toroidal candidates identified in Ref.~\cite{bra24} are summarized in
Fig.~\ref{fig:compall}. 
Low-energy transitions up to 11 MeV with toroidal content in the QRPA calculations described above have been compared to each experimental candidate in a $\chi^2$ fit procedure.
The best fits are shown as orange bands, where the size of the band indicates the total uncertainty from a Monte Carlo variation of the experimental uncertainties. 
Good agreement with the data is obtained in all cases.

{\it Discussion} -- Some important points addressed in the Supplemental Material~\cite{SM} are briefly sketched. 
It is known that single-particle spectra significantly depend on the IS effective mass $m^*/m$: the lower $m^*/m$, the more stretched is the spectrum \cite{IJMPE08}. 
In this connection, the dependence of QRPA results on $m^*/m$ is inspected using a representative set of Skyrme forces~\cite{IJMPE08}.
It is shown that the force SV-mas10 ($m^*/m=1.0$) provides a satisfactory description of both the low-energy spectrum of IS $E1$ strength \cite{poe92} and the $(e,e')$ data \cite{met87} in $^{58}$Ni.
Further, the role of the QRPA residual interaction is investigated and the microscopic structure of the toroidal $1^-$ states is analyzed.

For some low-energy toroidal states as well as representative GDR and compression states, the Coulomb ($F^C_{E1}$) and transverse ($F^T_{E1}$) form factors are calculated.  
At small momentum transfers $q_\mathrm{eff} \leq 0.6$ fm$^{-1}$ covered in the $(e,e')$ experiment \cite{met87}, the relative sign of the form  factors $\varPi={\rm sign} [F_{E1}^C/F_{E1}^T]$ is $+1$ for the toroidal states and $-1$ for both  GDR and compression mode. 
The sign can be measured through the $F^C_{E1} \cdot F^T_{E1}$ interference term in the $(e,e'\gamma)$ reaction \cite{pap85} and provides an independent experimental observable to discriminate vortical toroidal and irrotational GDR/compression modes.

In the present study, we focus on toroidal states located at $6 - 11$ MeV which coincides with the energy region of low-lying isovector dipole (IV1) strength, in nuclei with large neutron excess often called pygmy dipole resonance (PDR). 
There is debate about a possible collective nature of low-lying  IV1 strength in nuclei with neutron excess \cite{paa07,bra19,lan23,sav13,Reinhard2013b,mar24}.
Although our case $^{58}$Ni has $N\approx Z$, two observations connect the present results to this discussion.
The QRPA models successfully describing the toroidal mode in $^{58}$Ni also predict a toroidal nature of the PDR in neutron-rich nuclei \cite{rep19}.
Experimentally, the toroidal states identified in $^{58}$Ni exhibit the same features as the low-lying low-lying $1^-$ states in neutron-rich nuclei forming the PDR, viz.\ large g.s.\ branching ratios, large $B(E1)$ strengths with respect to average low-energy IV strengths, and a strong IS response.
Since the TDR is a generic mode appearing in all nuclei, these findings suggest a genuine toroidal nature of PDR modes.
A measurement of transverse electron scattering form factors in a nucleus with large neutron excess would be an important step to resolve this long-standing issue.

{\it Conclusions} -- Using a combined analysis of high-resolution inelastic scattering experiments with photons, protons, and electrons as well as self-consistent QRPA calculations with Skyrme forces, we identify, for the first time, candidates for toroidal electric dipole low-energy states in the spherical nucleus $^{58}$Ni.
It is shown that the large transverse electron scattering form factors of these states represent a relevant experimental observable to prove their toroidal nature.
A good agreement of the QRPA results with $(e,e')$ data \cite{met87} is obtained and a new interpretation of these data in terms of toroidal excitations is proposed. 

The QRPA results also predict the relative sign of Coulomb and transverse $E1$ form factors reaction as an independent observable to distinguish toroidal from IS compression and IV $E1$ transitions.
The sign is experimentally accessible through the $F^C_{E1} \cdot F^T_{E1}$ interference term in the $(e,e^\prime\gamma)$ reaction.
A recently commissioned setup for $(e,e^\prime\gamma)$ studies \cite{ste22} at the S-DALINAC facility \cite{pie18} promises such a measurement for $^{58}$Ni but also for low-energy $E1$ transitions in heavier nuclei with neutron excess.

As compared with other quantum systems, atomic nuclei suggest unique possibilities to discriminate intrinsic vortical excitations from other electric dipole modes utilizing the selectivity of nuclear reactions and high-resolution experiments to resolve individual states.
However, experimental methods to observe and identify vortical intrinsic excitations in nuclei are still poorly developed. Our study is one of the first steps in this direction. 
We hope that it will initiate progress in this fascinating topic.

This work was supported by the Deutsche Forschungsgemeinschaft (DFG) under Contract SFB 1245 (Project
ID No.~79384907). 
JK appreciates the support by a grant of the Czech Science Agency No 19-14048S.
ARe acknowledges support by the Slovak Research and Development Agency under Contract No. APVV-20-0532 and by the Slovak grant agency VEGA (Contract No. 2/0175/24). MS acknowledges financial support by UK-STFC (grant ST/P005101/1).

\clearpage

\section*{Supplemental Material}

{\bf Experimental data and analysis} -
The $(p,p^\prime)$ experiment and the comparison with  $(\gamma,\gamma^\prime)$, and $(e,e^\prime)$ results permitting the identification of toroidal candidates are presented in Ref.~\cite{bra24}.  
The electron scattering data shown in Figs.~3 and 4 of the main paper were taken from Ref.~\cite{met84} and are provided in Table \ref{tab-1}. 
The total uncertainties include besides the quoted statistical errors a 17\% systematic error \cite{met84} added in quadrature.
No $(e,e^\prime)$ data at backward angles are available for the 6.03 MeV candidate.

\begin{table}[b]
\caption{Experimental cross sections of the toroidal candidates in the $^{58}$Ni$(e,e^\prime)$ reaction \cite{met84}. 
Shown are excitation energies, scattering angles, cross sections normalized to Mott cross section and statistical error $\Delta$.} 
\vspace{12pt}
\label{tab-1}
    \begin{tabular}{|c|c|c|c|c|c|c|c|}
        \hline
        $E_x$ & $\theta$ & $\sigma/\sigma_\text{Mott}$ & $\Delta$ & 
        $E_x$ & $\theta$ & $\sigma/\sigma_\text{Mott}$ & $\Delta$ \\
        \ [MeV] & [deg]& $ \times 10^{-6}$ & [\%] &
        [MeV] & [deg]& $ \times 10^{-6}$ & [\%] \\
        \hline
              &  92.9 & 7.94 & 5.2 &
              &  92.9 &  6.05 & 5.4\\
              &  92.9 & 9.53 & 5.0 &
              &  92.9 &  6.78 & 6.9\\
        8.240 & 116.9 & 14.3 & 3.9 &
        8.516 & 116.9 & 11.8 & 4.6\\
              & 140.9 & 31.3 & 7.5 &
              & 140.9 & 26.9 & 7.7\\
              & 164.9 & 112 & 6.1 &
              & 164.9 & 95.0 & 7.1\\
            \hline
              &  92.9 & 2.61 & 16 &
              &  92.9 & 1.03 & 35\\
              &  92.9 & 3.67 & 14 &
              &  92.9 & 2.82 & 19\\
        8.875 & 116.9 & 5.00 & 12 &
        9.368 & 116.9 &  2.82 & 19\\
              & 140.9 & 10.4 & 22 &
              & 140.9 &  6.39 & 16\\
              & 164.9 & 63.5 & 11.5 &
              & 164.9 &  59.0 & 8.6\\ 
              & & & &
              & 164.9 &  37.05 & 17\\             
        \hline
              & 92.9  & 1.65 & 32 &
              & 92.9  & 2.37 & 16 \\             
              & 116.9 & 0.98 & 54 &
              & 116.9 & 5.05 & 11.2 \\    
        9.513 & 140.9 & 5.00 & 21 &
       10.036 & 140.9 & 13.7 & 11.0 \\        
              & 164.9 & 37.6 & 13 &
              & 164.9 & 46.1 & 10.5 \\                   & 164.9 & 21.6 & 29 &
              & & & \\  
        \hline
               & 92.9  & 2.20 & 18 &
               & 92.9  & 1.89 & 20\\             
               & 116.9 & 3.90 & 15 &
               & 116.9 & 3.98 & 14\\             
        10.073 & 140.9 & 9.18 & 18 &
        10.514 & 140.9 & 11.0 & 14\\        
               & 164.9 & 38.1 & 13 &
               & 164.9 & 52.8 & 10.8\\             
        \hline
    \end{tabular}
\end{table}

The comparison with QRPA results shown in Fig.~4 of the main paper was performed as follows: 
All QRPA states corresponding to $E1$ transitions in the excitation energy region up to 11 MeV were compared to all experimental angular distributions of toroidal candidates in a one parameter, error-weighted fit analogous to the DWBA analysis described in Ref.~\cite{met87}.
The results with the smallest $\chi^2$ values are displayed.
The normalization factors vary between 0.15 and 2.1.

{\bf Skyrme QRPA} -
Our Skyrme-Hartree-Fock model  employs an effective energy-density functional with Skyrme forces, for
a review see \citep{ben03}. 
Pairing is treated in the Bardeen-Cooper-Schriefer (BCS) approximation \cite{ben00,rep17}. 
The one-phonon excitations are obtained within the QRPA method with Skyrme forces \cite{rep15,rep15a}.
The method is fully self-consistent because ground state and excitations are derived from the initial Skyrme functional. 
Both particle-hole and particle-particle channels are taken into account \cite{rep15,rep17}.
Spurious admixtures from the center-of-mass motion are safely removed \cite{rep19}.
A large configuration space is used such that the Thomas-Reiche-Kuhn \cite{rin80} and isoscalar dipole
\cite{har01} energy-weighted sum rules are fully exhausted.

{\bf Transition strengths} -
An important property characterizing excited states are transition strengths. In Fig.~2 of the main text,
they are shown in terms of reduced transition probabilities \cite{rep13,kva11,nes18}
\begin{eqnarray}
B(\rm{x},IS1,\nu)&=&\sum_{\mu =-1}^{+1}|\:\langle\nu|\:\hat{M}_\mathrm{x}(\;IS1\mu)\:|0\rangle \:|^2 ,
\label{BIS1}
\\
B(E1,\nu)&=&\sum_{\mu = -1}^{+1}|\:\langle\nu|\:\hat{M}(\;E1\mu)\:|0\rangle \:|^2.
\label{BE1}
\end{eqnarray}
Here $\text{x=tor,com}$; $|0\rangle$ and $|\nu\rangle$ are QRPA
ground and excited $\nu$-th dipole states. The matrix elements read \cite{kva11,rep13,nes18}
\begin{equation}
\label{E1IV}
\langle\nu|\hat{M}(E1\mu)|0\rangle
= e \sum_{k =p,n} e^{\text{IV,k}}_{\text{eff}}
\int  d^3r r Y_{1\mu} \delta \rho^{\nu}_{k}(\mathbf{r}) \; ,
\end{equation}
\begin{eqnarray}
\nonumber
&&\langle\nu|\hat{M}_{\text{tor}}(IS1\mu)|0\rangle
= -\frac{e}{10 \sqrt{2}c}
\\
\label{TM_curl}
&& \;\;\;\;\;\;\;\; \cdot \int d^3r \{ r [r^2+d^s]
{\bf Y}_{11\mu} \cdot ( \vec{\nabla} \! \times \!
\delta\bf{j}_{c}^{\nu} (\bf{r})) \} ,
\\
\nonumber
&&\langle\nu|\hat{M}_{\text{com}}(IS1\mu)|0\rangle
=  -\frac{ie}{10c}
\\
&& \;\;\;\;\;\;\;\;  \cdot \int d^3r \{ r[r^2+d^s]
Y_{1\mu} (\bf{\nabla} \cdot \delta \bf{j}^{\nu}_c (\bf{r})) \} ,
\label{CM_div}
\end{eqnarray}
where $e^{\text{IV,p}}_{\text{eff}}=N/A$ and $e^{\text{IV,n}}_{\text{eff}}=-Z/A$
are proton and neutron effective charges;
$Z, N$, and $A$ are the proton, neutron and mass number, respectively;
$\bf{Y}_{11\mu}(\hat{\bf r})$ and $Y_{1\mu}(\hat{\bf r})$ are vector and ordinary spherical harmonics; $d^s= - 5/3 \langle r^2\rangle_0$ is the center-of-mass correction \cite{rep19}, where $\langle r^2\rangle_0=\int\:d^3r r^2 \:\rho_0 /A$, and $\rho_0(r)$ is the g.s.~density; $\delta\rho^{\nu}_k(\vec{r})=\langle\nu|\hat{\rho}_k|0\rangle(\vec{r})$
is the proton/neutron transition density (TD) from ground state to the state $\nu$ and
$\delta\bf{j}^{\nu}_c(\bf{r})=\langle\nu|\hat{\bf{j}}_c|0\rangle(\bf{r})$
is the total (proton + neutron) convective  part of the current transition density (CTD).
The magnetic part of the current does not contribute significantly to IS toroidal and compression modes \cite{kva11}, thus the matrix elements (\ref{TM_curl},\ref{CM_div}) include only the convective contribution. 
The proton and neutron parts of the convective CTD are illustrated in the r.h.s.\ of Fig.~2 in the main text.

The toroidal matrix element (\ref{TM_curl}) employs the curl $(\bf{\nabla} \! \times \! \delta \bf{j}^{\nu}(\bf{r}))$ and thus is basically vortical.
The compression matrix element (\ref{CM_div}),
accesses the divergence $(\bf{\nabla} \cdot \delta
\bf{j}^{\nu}(\bf{r}))$ and thus is mainly irrotational.

If the transition strength is considered in a large energy range and/or we deal with a
high density of states, it is convenient to use strength functions
\begin{equation}
\label{sf}
S({\rm x}; E) = \sum_{\nu} B(x)_{\nu} \: \xi_{\Delta}(E-E_{\nu})
\end{equation}
where $E$ is the excitation energy and
\begin{equation}
\xi_{\Delta}(E-E_{\nu}) = \frac{1}{2 \pi}
\frac{\Delta}{(E-E_{\nu})^2 + (\Delta/2)^2}
\label{lw}
\end{equation}
is the Lorentz weight with the averaging parameter $\Delta$. \\

\begin{figure*} 
\centering
\includegraphics[width=12cm]{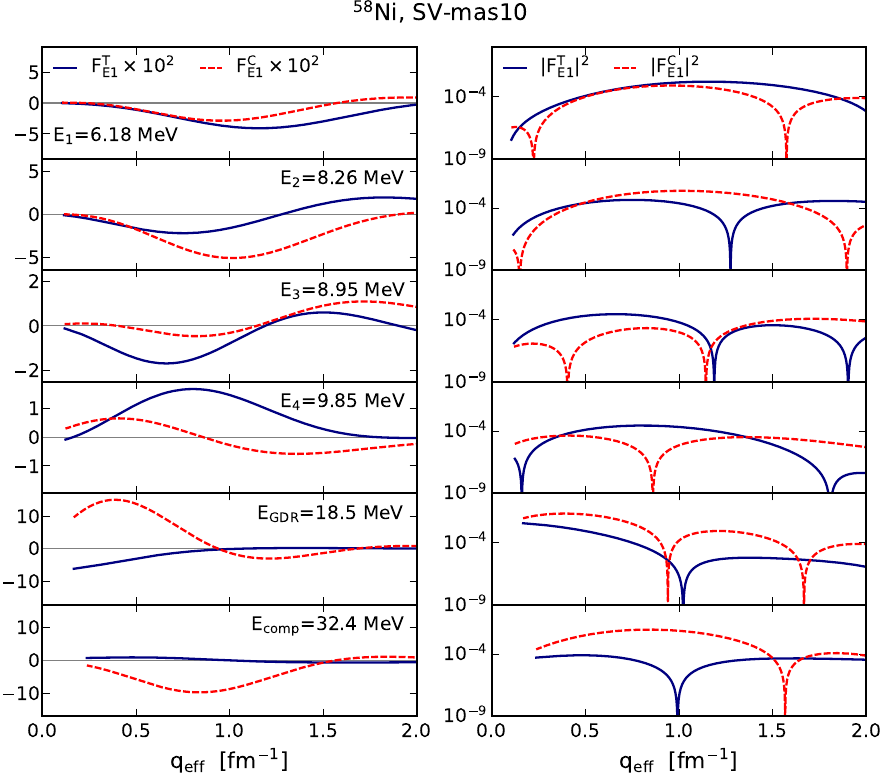}
\caption{
{\normalfont SV-mas10 Coulomb $F_{E1}^C$ (dashed red line) and transversal  $F_{E1}^T$ (solid black line)
form factors (left) and their squared values (right) for the transitions to the first four $1^-$ QRPA states,
to the most prominent GDR state at 18.5 MeV and to the strongest high-energy compression state at 32.4 MeV in $^{58}$Ni.
The horizontal lines in the left panels separate positive and negative values of the form factors.
Results are calculated for a fixed scattering angle ($\theta=178.5^\circ$).
}}
\label{fig1:FC+FT}
\end{figure*}

{\bf Electron scattering cross sections and form factors} --
Information on the TD and CTD can be gained from the differential inelastic $(e,e')$ cross section. 
For $E\lambda$ transitions in plane-wave Born approximation (PWBA) it reads \cite{HB83}
\begin{eqnarray}
\label{PWBAcs}
&&\frac{d\sigma}{d\Omega} (\theta, q_\mathrm{eff}, E_i) \simeq
\frac{4\pi}{Z^2} \sigma_\mathrm{Mott}(\theta, E_i) f_\mathrm{rec}(\theta, E_i)
\\
&\times& \left[\left|F^C_{E\lambda} (q_\mathrm{eff})\right|^2
+ \left(\frac{1}{2}+\tan^2(\frac{\theta}{2})\right)
\left|F^T_{E\lambda} (q_\mathrm{eff})\right|^2  \right]
\nonumber
\end{eqnarray}
where
\begin{equation}
\sigma_\mathrm{Mott} (\theta, E_i)=
 \left[
\frac{e^2Z}{8\pi E_i}  \frac{\cos(\frac{\theta}{2})}{\sin^2 (\frac{\theta}{2})}
\right]^2
\label{sigm}
\end{equation}
is the Mott cross section for a point-like object with charge $Z$, $E_i$ is the incident electron energy, and $\theta$ is the scattering angle. 
In the present study with $E_i \approx $ 50 MeV and
large scattering angles, the recoil factor is neglected, i.e.\ we use $f_\mathrm{rec}$=1. 
Note that the transverse form factor \cite{HB83}
\begin{equation}
 F^T_{E\lambda} (q) \sim \int dr^3 
(\nabla \times \delta{\bf j}^{\nu}) \cdot j_{\lambda}(kr) {\bf Y}_{\lambda\lambda\mu} 
\end{equation}
includes curl of CTD and so prefers more a vortical flow than an irrotational one.
For excitations in spherical even-even nuclei, the Coulomb and transverse form factors read \cite{HB83}
\begin{eqnarray}
\label{FC}
F^C_{E\lambda}(q) &=& 
\sqrt{2\lambda +1} \int_0^{\infty} dr r^2
 \delta \rho_{\lambda}^{\nu}(r) j_{\lambda}(qr)
\\
F^T_{E\lambda} (q)&=& 
\frac{1}{c} \int_0^{\infty} dr r^2
[\sqrt{\lambda +1} \; \delta J^{\nu}_{\lambda, \lambda -1}(r) j_{\lambda-1}(qr)
\nonumber
\\
\;\;\;  &-&\sqrt{\lambda} \; \delta J^{\nu}_{\lambda, \lambda +1}(r) j_{\lambda+1}(qr)]
\label{FT}
\end{eqnarray}
where  $\delta J^{\nu}_{\lambda, \lambda \pm 1}(r)$
are radial components of the full (convective + spin) CTD
$\delta\vec{j}^{\nu}=\delta\vec{j}^{\nu}_c+\delta\vec{j}^{\nu}_m$.

\begin{figure*} 
\centering
\includegraphics[width=16cm]{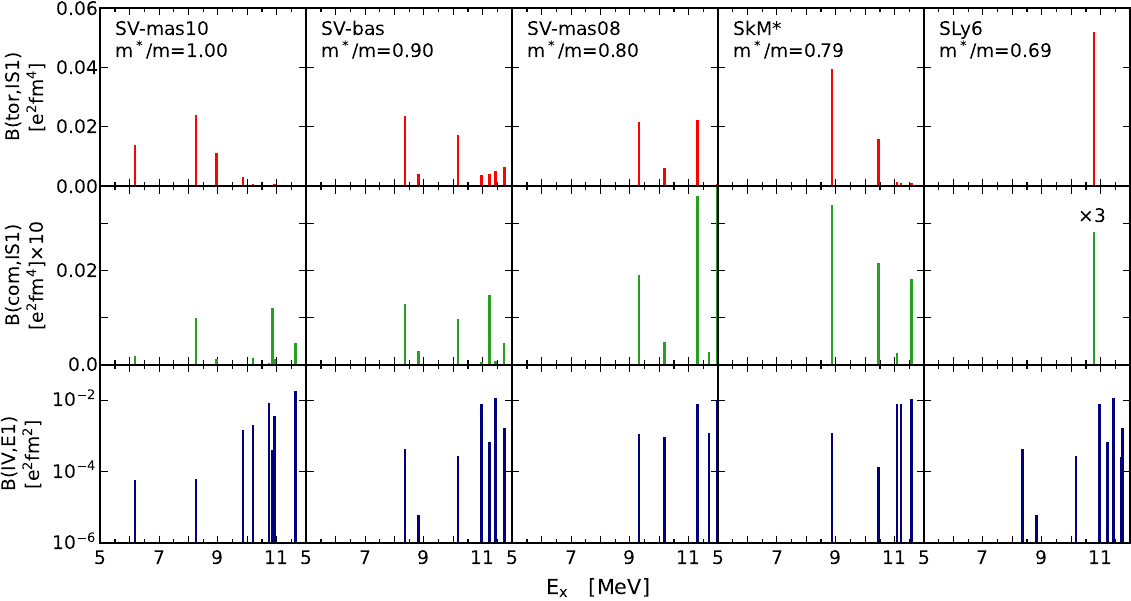}
\caption{
{\normalfont Effective mass dependence of the IS1 toroidal (top), IS1 compression (middle) and IV E1 (bottom) strengths in $^{58}$Ni in
the energy region 5-12 MeV, calculated  with forces SVmas-10, SV-bas (=SVmas-09), SVmas-08, SkM* and SLy6.
The corresponding effective masses values are given in the figure. For better visibility, the compression
strength is multiplied by a factor of 10 and the IV E1 strength is shown on a logarithmic scale.}}
\label{fig2:B(E1)}
\end{figure*}

The effective transfer momentum
\begin{equation}
 q_\mathrm{eff}= q\left(1+1.5 \frac{Z \alpha \hbar c}{E_i R}\right)
 \label{qeff}
\end{equation}
with
\begin{equation}
 q =\frac{2}{\hbar c}\sqrt{E_i E_f} \sin \left(\frac{\theta}{2}\right)
 \label{q}
\end{equation}
is used to roughly take into account Coulomb distortions.
Here, $E_f=E_i-E_{\nu}$ is the final electron energy, $E_{\nu}$ is the nuclear excitation energy, and $R=1.2 A^{1/3}$ fm is the nuclear radius.

Figure \ref{fig1:FC+FT} shows the Coulomb $F_{E1}^C$ and transverse $F_{E1}^T$ form factors (FF) for the four lowest states computed with SV-mas10. 
For comparison, the same FF for the transitions to the GDR state at 18.5 MeV (with the largest $B(\rm{IV,E1})$ strength) and the compression $1^-$-state at 32.4 MeV (with the largest $B(\rm{com,IS1})$ strength) are also shown.
At the lowest $q_\mathrm{eff}$, all the states have similar values for Coulomb and transverse FF, i.e.
$F_{E1}^C \approx F_{E1}^T$. 
However, the FF exhibit a different evolution with $q_\mathrm{eff}$.
For the states at 6.18 and 8.26 MeV, $F_{E1}^C$ and $F_{E1}^T$ in the range 0.2 fm$^{-1} < q_\mathrm{eff} < $ 0.6 fm$^{-1}$ acquire similar absolute values and the same sign, i.e. $\varPi={\rm sign} [F_{E1}^C/F_{E1}^T] = +1$.
For the states at 8.95 and 9.85 MeV, also $\varPi = +1$ but $F_{E1}^T$ becomes much larger than $F_{E1}^C$.
Instead, for GDR and compression states, we get $|F_{E1}^C| \gg |F_{E1}^T|$ and both FF have opposite signs.

These results can be easily explained. 
In the limit $q_\mathrm{eff} \to 0$ (Siegert theorem), we approach the ratio $F_{E1}^C/F_{E1}^T \to - (\sqrt{2}\hbar q c)/E_{\nu}$.
This limit is relevant for dominant dipole transitions, and indeed for the GDR transition we find the proper sign $\varPi=-1$. 
At momentum transfers 0.2 fm$^{-1} < q_\mathrm{eff} < $ 0.6 fm$^{-1}$, the toroidal transitions give the opposite sign  $\varPi$= +1.
The toroidal mode is generated by the second term in the  expansion of the spherical Bessel function
 \begin{equation}
j_{\lambda}(qr) = \frac{(qr)^{\lambda}}{(2 \lambda +1)!!} \:
\left[\:1 - \frac{(qr)^2}{2(2 \lambda +3)} + \ldots \: \right]
\label{18}
\end{equation}
 entering the electric transition operator and form factors.
It is seen that this term has the opposite sign 
as compared with the leading one. The second-order compression term also has the opposite sign. However, this  irrotational mode should exhibit a very suppressed transverse form factor, as verified in our calculations.  
The relative sign $\varPi$ of the Coulomb and transverse FF can be measured in the $(e,e'\gamma)$ reaction \cite{pap85} and used for additional discrimination of the vortical toroidal and irrotational GDR/compression excitations.

The right panel of Fig.~\ref{fig1:FC+FT} shows the squared Coulomb and transverse form factors at $0-2$~fm$^{-1}$.
Their ratio can be strongly affected by local minima. 
In particular, in the range $0.2-0.6$~fm$^{-1}$ of our main interest, the behavior of the form factors for the states with $E_{\nu} < 9$ MeV is affected by the first minimum of $|F^C_{E1}|^2$, whose precise position is state-dependent.

{\bf Impact of effective mass and residual interaction} -
The QRPA results significantly depend on the isoscalar effective mass $m^*/m$ of the applied Skyrme force \cite{nest08}. 
It is known that the smaller $m^*/m$, the more stretched is the single-particle spectrum and the higher the 
energies of the lowest QRPA states. 
Thus, it is necessary to explore the impact of $m^*/m$ in more detail.

Figure~\ref{fig2:B(E1)} shows QRPA toroidal, compression and IV strengths of low-energy dipole excitations
in the region $5-12$ MeV for a representative set of five Skyrme forces (three SV forces \cite{SV} (SV-mas10,
SV-bas, SV-mas08) and two alternative forces SkM* \cite{SkMs} and SLy6 \cite{SLy6}) with different isoscalar
effective masses $m^*/m$.  SV forces were fitted within the same protocol and therefore allow a refined
estimation of the impact of $m^*/m$. Inclusion of SkM* (with an effective mass comparable to SV-mas08) and SLy6
additionally permits to test the relevance of the interaction. 

As visible in Fig.~\ref{fig2:B(E1)}, the spectra depend quite dramatically on $m^*/m$. SV-mas10 ($m^*/m=1$) provides the best description of the lowest experimentally observed states and was therefore chosen for further analysis.
Note that SV-mas10 was earlier successfully applied for a description of the isoscalar giant quadrupole resonance \cite{KuPLB18}.
Effective mass values $m^*/m < 1$ lead  to the significant energy upshift of the toroidal strength.
SV-bas, SV-mas08, SkM$^*$ ($m^*/m=0.79$) and SLy6 ($m^*/m=0.69$) do not give any dipole states below 8.5, 9.2, 8.8 and 10.8 MeV, respectively. The figure also demonstrates that at low energies, independent of the chosen interaction, the vortical toroidal response always strongly dominates over the irrotational compressional one (note that the latter is multiplied by a factor of 10 in Fig.~\ref{fig2:B(E1)}).
Note that this dominance takes place not only in $^{58}$Ni but also in many other spherical nuclei
\cite{rep19,rep13}. 
At the same time, in some particular nuclei, individual states with significant compression fraction are also predicted at low energies \cite{rep19,nes18}.

\begin{figure}
		\centering
		\includegraphics[width=0.85\columnwidth]{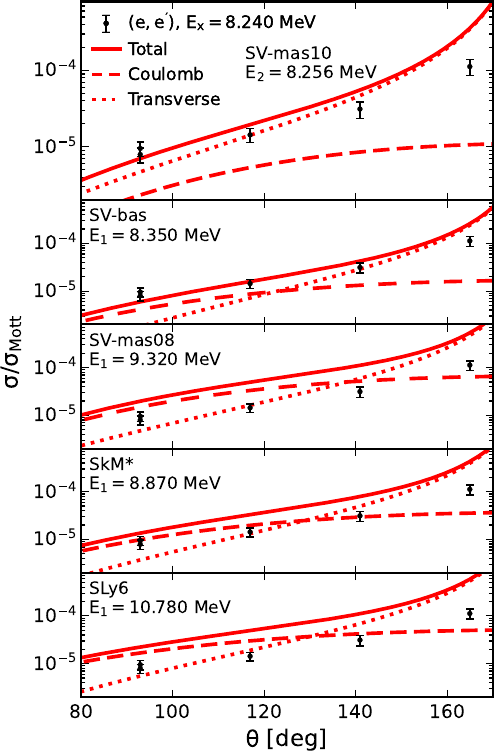}
\caption{Electron  scattering cross sections of the toroidal experimental candidate at 8.240 MeV compared with SVmas-10, SV-bas, SV-mas08, SkM* and SLy6 predictions.
The QRPA energies $E_{\nu}$ are indicated.}
	\label{fig:ssm}
	\end{figure}

In Fig.~\ref{fig:ssm}, the dependence of the electron scattering cross sections on the different
Skyrme parametrizations  is investigated using (like in Fig.~3 of the main text) the
strongest experimental transition at 8.24 MeV \cite{met87} as an example.
We consider the lowest $1^-$ states for SV-bas, SV-mas08, SkM* and SLy6 and the second
$1^-$ state for SV-mas10, as the most relevant counterparts.
A superior agreement with experiment is obtained for SV-mas10 ($m^*/m$=1) and SV-bas ($m^*/m$=0.9),
in particular describing the strong increase towards larger angles. The description gets worse with
further decreasing effective mass. The angular dependence of the transverse part is similar for all
Skyrme parametrizations and its magnitude varies by less than a factor of 2 indicating that large
transverse form factors are a general prediction independent of the specific interaction.
The main difference lies in the Coulomb form factor, whose magnitude is related to the collectivity
of the transition.

\begin{figure} 
\centering
\includegraphics[width=\columnwidth]{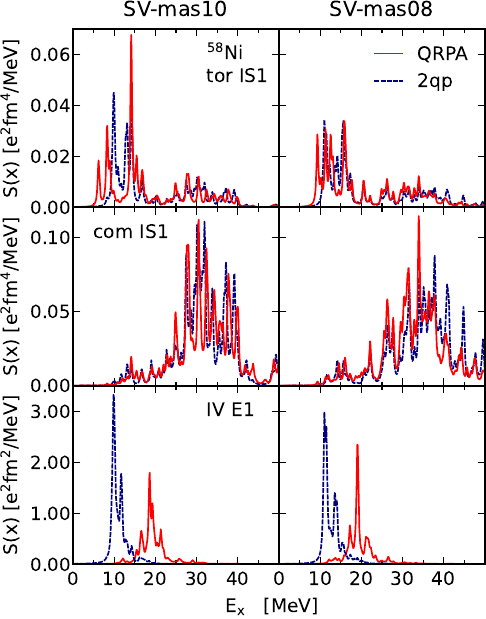}
\caption{Toroidal isoscalar (x=tor IS1), compression isoscalar (x=com IS1) and
E1 isovector (x=IV E1) 2qp (dotted blue line) and QRPA (dashed red line)
dipole strength functions, calculated with the forces SV-mas10 (left) and SV-mas08 (right).}
\label{fig4:meff}
\end{figure}

\begin{table}[t] 
\renewcommand{\arraystretch}{1.0}
\caption{Characteristics of toroidal QRPA $1^-$ states  in $^{58}$Ni calculated with the SV-mas10 interaction. For each state, we
show the excitation energy $E$, toroidal reduced transition probability  B(tor, IS1), the main four
2qp $(nlj)_q(nlj)_{q'}$ proton (pp) or neutron (nn) components with their 2qp energies $\epsilon_{qq'}$
and contributions $N_{qq'}$ to the state norm.}
\label{tab-2}
\begin{center}
\begin{tabular}{|c|c|c|c|c|}
\hline
$E$ & B(tor,IS1) & \multicolumn{3}{|c|}{main 2qp components $qq'$}
\\
\hline
 [MeV] & $\rm [e^2 fm^4 10^{-3}]$ & $(nlj)_q(nlj)_{q'}$
      &  $N_{qq'} [\%]$  &  $\epsilon_{qq'}$ [MeV]\\
\hline
  6.18 & 14.0 &  pp $2p_{3/2}, 1d_{3/2}$ &  42 & 8.2 \\
       &     &  pp $2p_{1/2}, 2s_{1/2}$ & 11 & 11.0 \\
       &     &  nn $2p_{3/2}, 1d_{3/2}$ &  10 & 9.8 \\
       &     &  nn $2p_{1/2}, 2s_{1/2}$ & 9 & 11.6 \\
\hline
  8.26 & 24.0 &  pp $2p_{1/2}, 2s_{1/2}$ &  22 & 11.0 \\
       &      & nn $2p_{1/2}, 2s_{1/2}$ &  15 &  11.6 \\
        &     &  pp $1g_{9/2}, 1f_{7/2}$ & 10 & 9.7 \\
       &      &  pp $2p_{3/2}, 2s_{1/2}$ & 8 & 9.4 \\
\hline
 8.95 & 11.2 &  pp $2p_{3/2}, 1d_{3/2}$ &  48 & 8.2 \\
       &    &   pp $2p_{1/2}, 1d_{3/2}$ &  23 &  9.9 \\
      &     &   nn $1g_{9/2}, 1f_{7/2}$ & 6 & 10.0 \\
      &     &  pp $2p_{1/2}, 2s_{1/2}$ & 5 & 11.0 \\
\hline
  9.85 & 3.2 &  nn $2p_{3/2}, 1d_{3/2}$ &  76 & 9.8 \\
       &     &  pp $1g_{9/2}, 1f_{7/2}$ &  5 &  9.7 \\
       &     &  pp $2p_{1/2}, 2s_{1/2}$ & 4 & 11.0 \\
      &     &   pp $1f_{5/2}, 1d_{3/2}$ & 4 & 11.6 \\
\hline
  10.19 & 0.69 &  pp $2p_{1/2}, 1d_{3/2}$ &  29 & 9.9 \\
        &     & pp $2p_{3/2}, 2s_{1/2}$ &  22 & 9.4 \\
        &     &   nn $1f_{5/2}, 1d_{3/2}$ & 13 & 11.8 \\
         &     &  nn $1g_{9/2}, 1f_{7/2}$ &  9 &  10.0 \\
\hline
  14.11 & 52.3 &  pp $2d_{5/2}, 1f_{7/2}$ &  34 & 13.1 \\
        &       & pp $2p_{3/2}, 1d_{5/2}$ &  22 & 12.7 \\
         &      &  pp $1f_{5/2}, 1d_{3/2}$ & 17 & 11.6 \\
        &       &  nn $2d_{5/2}, 1f_{7/2}$ &  12 & 14.2 \\

\hline
\end{tabular}
\end{center}
\end{table}

An additional illustration of the influence on the effective mass $m^*/m$ is provided in
Fig.~\ref{fig4:meff} where unperturbed two-quasiparticle (2qp) and QRPA IS1 toroidal, IS1
compression and IV E1 strength functions (\ref{sf}) calculated  with SV-mas10 and SV-mas08
are compared. The strength functions are smoothed with a Lorentzian of width $\Delta = 0.5$~MeV
to improve the visibility.
Figure~\ref{fig4:meff} shows that both 2qp and QRPA strengths are somewhat upshifted in energy
with decreasing $m^*/m$.
A similar trend was earlier demonstrated  for the QRPA case in Fig.~\ref{fig2:B(E1)}.
The origin of the trend is clear: the smaller $m^*/m$, the more stretched is the single-particle
spectrum  \cite{nest08}. However, the IV residual interaction,
being strongly repulsive, smoothes this difference \cite{spe20} and finally sets the calculated IV
giant dipole resonance for both forces to an energy of about 19 MeV in agreement with expectations
from experimental systematics \cite{har01}.

For the compression IS1 mode with its wide distribution, the impact of the residual interaction is modest.
This can be explained by the different radial dependence of the compression mode ($\sim r^3$) and main dipole
residual interaction ($\sim r$). The toroidal IS1 mode exhibits appreciable strength at $8-20$ MeV already
in the 2qp case. This justifies the notion of a mean-field origin of the toroidal mode.
For SVmas-10, this mode is noticeably affected by both IS and IV residual interactions.
The IS interaction downshifts the toroidal strength and creates rather collective toroidal peaks at 6 and 8 MeV.
Instead, the IV interaction upshifts the toroidal strength and forms the strong peak at 14 MeV.
For SV-mas08, the impact of the residual interaction for the strength below 10 MeV is less pronounced.

Some features of the main toroidal QRPA states calculated with SV-mas10 are shown in Table \ref{tab-2}.
All these states, except the 9.85-MeV state, are rather collective: a few strong 2qp states cooperate to
the extent that the contribution of the main $2qp$ component does not exceed 48$\%$, while the sum of
the four largest components reaches 55-85$\%$. The rest consists of many small further
2qp components. Following Table \ref{tab-2}, the energy downshift $\Delta E =\epsilon^{\rm min}_{qq'}-E_1$ between
the lowest  2qp and QRPA states is typically 2 MeV confirming the important role of the IS residual interaction.
All QRPA states in Table \ref{tab-2} contain large 2qp components $[2p_{3/2}, 1d_{3/2}]$, $[2p_{3/2}, 1d_{5/2}]$, $[2p_{1/2}, 1d_{3/2}]$, $[2p_{3/2}, 2s_{1/2}]$, $[2p_{1/2}, 2s_{1/2}]$ and $[2d_{5/2}, 1f_{7/2}]$.
As was checked by our calculations, these 2qp components produce the vortical flow resembling the toroidal pattern.
This once again points toward the mean-field origin of the toroidal flow.

\end{document}